%% file: main.tex
\documentclass[conference]{IEEEtran} 

\usepackage{etex}

\def\papertitle{Successive-Cancellation Flip Decoding of Polar Codes Under Fixed Channel-Production Rate}

\usepackage{ifpdf}
\ifpdf
\usepackage[draft,pdfborder={0 0 0}]{hyperref}

\fi

\usepackage{cite}
\usepackage{graphicx}
\usepackage{hyperref}

\usepackage{epstopdf}

\usepackage{amssymb}
\usepackage[cmex10]{amsmath}
\usepackage[subnum]{cases}
\usepackage{nicefrac}
\usepackage{bm} 

\usepackage{algorithm}
\usepackage{algorithmicx}
\usepackage{algpseudocode}

\usepackage{glossaries}

\usepackage{tabularx}
\usepackage{multirow}
\usepackage{dcolumn}
\usepackage{booktabs}

\usepackage{tikz,pgf}
\usepackage{pgfplots}
\pgfplotsset{compat=1.16}
\usetikzlibrary{shapes}
\usetikzlibrary{arrows}
\usetikzlibrary{plotmarks,spy,calc}

\usepackage{balance}
\usepackage{caption}

\usepackage[markup=underlined]{changes}
\definechangesauthor[color=blue]{PG}

\newcommand{\fixme}[2]{\ifx&#2&{\color{red}#1}\else{\color{red}FIXME\{}#1{\color{red}\}}\footnote{{\color{red}#2}}\PackageWarning{Fixme}{#1: #2}\fi}

\title{\papertitle}

\author{
  \IEEEauthorblockN{Ilshat Sagitov, Charles Pillet and and Pascal Giard}
  \IEEEauthorblockA{
		Department of Electrical Engineering, \'Ecole de technologie sup\'erieure (\'ETS), Montr\'eal, Canada\\
    Email: \{ilshat.sagitov.1, charles.pillet.1\}@ens.etsmtl.ca, pascal.giard@etsmtl.ca
  }
}

\algnewcommand{\Inputs}[1]{%
  \State \textbf{Inputs:}
  \Statex \hspace*{\algorithmicindent}\parbox[t]{.8\linewidth}{\raggedright #1}
}

\algnewcommand{\Initialize}[1]{%
  \State \textbf{Initialize:}
  \Statex \hspace*{\algorithmicindent}\parbox[t]{.8\linewidth}{\raggedright #1}
}

\definecolor{MyBlue}{rgb}{0.00000,0.44700,0.74100}%

\newcommand{\plotfigureheight}{0.6}

\makeglossaries

\begin{document}
\bstctlcite{IEEEexample:BSTcontrol}

\newacronym{awgn}{AWGN}{additive white Gaussian noise}
\newacronym{snr}{SNR}{signal-to-noise ratio}
\newacronym{fer}{FER}{frame-error rate}
\newacronym{sc}{SC}{successive-cancellation}
\newacronym{scl}{SCL}{successive-cancellation list}
\newacronym{scf}{SCF}{successive-cancellation flip}
\newacronym{dscf}{DSCF}{dynamic \gls{scf}}
\newacronym{crc}{CRC}{cyclic-redundancy check}
\newacronym{llr}{LLR}{log-likelihood ratio}

\maketitle

\begin{abstract}
Polar codes are a class of error-correcting codes that provably achieve the capacity of practical channels under the low-complexity \gls{scf} decoding algorithm. 
However, the \gls{scf} decoding algorithm has a variable execution time with a high (worst-case) decoding latency. This characteristic poses a challenge to the design of receivers that have to operate at fixed data rates. In this work, we propose a multi-threshold mechanism that restrains the delay of a \gls{scf} decoder depending on the state of the buffer to avoid overflow. We show that the proposed mechanism provides better error-correction performance compared to a straightforward codeword-dropping mechanism at the cost of a small increase in complexity. In the region of interest for wireless communications, the proposed mechanism can prevent buffer overflow while operating with a fixed channel-production rate that is 1.125 times lower than the rate associated to a single decoding trial.

\end{abstract}

\section{Introduction}
\label{sec:intro}
Polar codes~\cite{arik_polariz} are a type of linear error-correction codes that can achieve the channel capacity for practically relevant channels under \gls{sc} decoding. However, at short to moderate block lengths, the \gls{sc} algorithm provides an error-correction performance that is lacking for many practical applications. To address this, the \gls{scl} decoding algorithm was proposed~\cite{scl_intro}. It provides great error-correction capability to the extent that polar codes were selected to protect the control channel in 3GPP's next-generation mobile-communication standard (5G), where \gls{scl} serves as the error-correction performance baseline~\cite{3GPP_5G_Coding}.
However, the great error-correction performance of a \gls{scl} decoder comes at the cost of high hardware implementation complexity and low energy efficiency~\cite{scl_5g}.

As an alternative to \gls{scl}, the \gls{scf} decoding algorithm was proposed~\cite{scf_intro}.
\Gls{scf} leads to an improved error-correction performance compared to \gls{sc}, but still falls short of that of an \gls{scl} decoder with a moderate list size. However, \gls{scf} is more efficient than \gls{scl} both in terms of computing resources and energy requirements~\cite{Giard_JETCAS_2017}.

\Gls{dscf} decoding was proposed in~\cite{dyn_scf}, where  modifications to \gls{scf} were made to improve error-correction performance. With these modifications, the error-correction performance approaches that of a \gls{scl} decoder with moderate list sizes at the cost of a minor increase of complexity compared to \gls{scf}.
Preliminary results from a hardware implementation indicate that \gls{dscf} decoders maintain a higher energy efficiency compared to \gls{scl} decoders~\cite{pract_dscf}.

Regardless of the variant, \gls{scf}-based decoders exhibit a variable execution time by nature, with a latency much higher than the average execution time. 
Some efforts were made to reduce the variability of the execution time~\cite{earl_stop_dscf}, but this characteristic cannot be fully eliminated. 
This poses a challenge to the realization of receivers that have to operate at fixed data rates.
To compensate for the variable execution time of the decoder, words arriving from transmitter with a fixed time interval have to be stored in a buffer. Without any additional mechanisms, a fixed-size buffer may overflow even under reasonable conditions, e.g., when the channel-production rate is only slightly slower than the average decoder throughput. 
To avoid overflow, one of the straight-forward approaches would be to drop the received words when the buffer approaches overflow, i.e., applying a codeword-dropping mechanism. However, a codeword-dropping mechanism severely affects the error-correction performance.

\subsubsection*{Contributions}

In this work, we present a system model for operation under fixed channel-production rate that notably includes a controller for a \gls{scf}-based decoder.
We propose a multi-threshold mechanism for that controller that modifies the maximum number of decoding trials by tracking the state of the input buffer. A codeword-dropping mechanism is used for reference. We provide a methodology for threshold selection. 
Simulation results are provided for various channel-production rates that are close to the rate associated to a single trial of \gls{scf} decoding.
They show that both codeword-dropping and multi-threshold mechanisms can operate at fixed channel-production rates and prevent buffer overflow. 
We show that the multi-threshold mechanism provides a better error-correction performance than the codeword-dropping approach. 

\subsubsection*{Outline} 

The remainder of this paper is organized as follows. \autoref{sec:backgr} provides a short introduction to polar codes and their construction, and briefly describes the \gls{sc} and \gls{scf} decoding algorithms. In \autoref{sec:syst_mod}, the system model is presented and the functionalities of each block of the model are described, with the exception of the controller. The controller is explained in \autoref{sec:contr_mech} along with the details on the codeword-dropping mechanism used for reference as well as the proposed multi-threshold mechanisms. In \autoref{sec:multithr_thr_sel}, the threshold-selection methodology for both mechanisms is provided. In \autoref{sec:results}, simulation results, in terms of the buffer-size variation and the error-correction performance, are provided and discussed. 
 \autoref{sec:conclusion} concludes this work.

\input{Background}

\input{SystemModel}

\input{ControlMechan}

\input{ThrSelect}

\input{Simul_Results}

\input{Conclusion}

\section*{Acknowledgement}
The authors thank Tannaz Kalatian for helpful discussions. Work supported by NSERC Discovery Grant \#651824.

\bibliographystyle{IEEEtran}
\bibliography{IEEEabrv,ConfAbrv,refs}

\end{document}

%% file: Background.tex
\section{Background}
\label{sec:backgr}
\subsection{Construction of Polar Codes}
The central concept of polar codes is channel polarization. As the code length tends to infinity, bit locations either become completely reliable or completely unreliable. To construct a $\mathcal{P}\left(N,k\right)$ polar code, where $N$ is the code length and $k$ the number of information bits, the $\left(N-k\right)$ least-reliable bits, called frozen bits, are set to predefined values, typically all zeros. The encoding is the linear transformation such that $\bm{x}=\bm{u}\times F^{\otimes n}$, where $\bm{x}$ is the polar-encoded row vector, $\bm{u}$ is a row vector of length $N$ that contains the $k$ information bits in their predefined locations as well as the frozen-bit values, $n=\log_2 N$, and $F^{ \otimes n}$ is the $n^{\text{th}}$ Kronecker product $\left( \otimes \right)$ of the binary polar-code kernel $F = \left[ \begin{smallmatrix} 1 & 0 \\ 1 & 1\end{smallmatrix} \right]$. The bit-location reliabilities depend on the channel type and conditions. In this work, the \gls{awgn} channel is considered and the construction method used is that of Tal and Vardy~\cite{tal_constr}. 

\subsection{Successive-Cancellation Decoding}
\Gls{sc} decoding is a natural way of decoding of polar codes as was introduced in the seminal paper~\cite{arik_polariz}. The received vector (channel \glspl{llr}), denoted by $\{\alpha_{\text{ch}}(0), \ldots , \alpha_{\text{ch}}(N-1)\}$, is used to estimate the bits of the polar-encoded word starting from the first bit $\hat{u}_0$~\cite{cam_schw}. The following bits $\{\hat{u}_1,\ldots, \hat{u}_{N-1}\}$ are estimated sequentially, i.e., in successive manner, by the same vector of channel \glspl{llr} and the estimations of the previous bits. Each information bit $\hat{u}_i$ is estimated by taking a hard decision on the corresponding decision \gls{llr}, denoted by $\alpha_{\text{dec}}(i)$. Frozen bits are known to the decoder and are thus directly set to their corresponding value, typically zero.

\subsection{SC-Flip Based Decoding}

The \gls{scf} decoding algorithm is introduced in~\cite{scf_intro}, where the authors observed that if the first erroneously-estimated bit could be detected and corrected before resuming \gls{sc} decoding, the error-correction capability of the decoder would be significantly improved. In order to detect the decoding failure of the codeword, the information bits are concatenated with a $r$-bit \gls{crc} being passed through the polar encoder. The code rate of the polar code is thus increased to $R=\left(k+r\right)/N$.

If the \gls{crc} check indicates decoding failure at the end of the initial \gls{sc} decoding pass, a list of bit-flipping candidates, denoted by $\mathcal{L}_{\text{flip}}$, is constructed. In the original \gls{scf} decoding algorithm, $\mathcal{L}_{\text{flip}}$ stores the bit indices that correspond to the non-frozen bits with the smallest absolute values $\alpha_{\text{dec}}$.
A more accurate metric for constructing $\mathcal{L}_{\text{flip}}$ is introduced in~\cite{dyn_scf}. This metric takes in account the successive nature of the decoder, and its calculation for each non-frozen bit with an index $i$ after the initial \gls{sc} attempt is defined as:
\begin{equation}
M_i = |\alpha_{\text{dec}}(i)| + \frac{1}{c} \cdot  \sum_{\substack{j \leq i \\ j\in \mathcal{A}}} \ln \left( 1 + e^{\left( -c \cdot |\alpha_{\text{dec}}(j)|\right)}\right)\,,
\label{eq:metr_dscf}
\end{equation}
where $\ln(\cdot)$ denotes the natural logarithm, $\mathcal{A}$ is the set of non-frozen bit indices, and  $c$ is a constant optimized experimentally by way of simulation. The value $c$ will vary in the range $0.0 < c\leq 1.0$ depending on polar code parameters and channel conditions.

Regardless of the type of metric, for each new decoding trial the next bit index of $\mathcal{L}_{\text{flip}}$ is selected and when this bit is estimated, the opposite decision is made, i.e., the estimated bit is flipped. Decoding then resumes until the last bit, following the \gls{sc} algorithm. New \gls{scf} trials are ran until the \gls{crc} matches or until the maximum number of trials $T_{\text{max}}$ is reached.
The maximum number of trials $T_{\text{max}} \in \mathbb{N}^+$ defines the decoding latency, and $1\leq T_{\text{max}} \leq \left( k+r\right)$. Setting $T_{\text{max}}$ to $1$ renders the \gls{scf} decoder equivalent to an \gls{sc} decoder. If after $T_{\text{max}}$ trials the \gls{crc} check fails, decoding is stopped and the word is considered undecodable.

We note that in~\cite{dyn_scf} the authors adapt the metric of \eqref{eq:metr_dscf} to allow multiple bit flips per trial and name the resulting algorithm \gls{dscf} decoding. Preliminary results of a hardware implementation of \gls{dscf} decoding~\cite{pract_dscf} show that a decoder that flips $2$ 
bits per trial is up to $5$ times more area-efficient compared to state-of-the-art \gls{scl} decoders while providing the same error-correction performance.
However, this comes at the cost of $11.5\%$ lower throughput compared to \gls{scl}. Without loss of generality, in this work, we do not apply multiple bit flips per trial. Thus, in the remainder of this work, the \gls{scf} decoder with metric calculation of \eqref{eq:metr_dscf} is applied.

\subsection{Execution time of \gls{scf}-based Decoders}

\Gls{scf}-based decoding algorithms have a variable execution time by nature. In this work, we assume that the latency of processing one decoding word under \gls{scf} is an integer multiple of the execution time of one \gls{sc} decoding pass. By denoting the latency of an \gls{sc} pass as $\tau_{\text{sc}}$, the execution time of one word under \gls{scf} decoding is calculated as:
\begin{equation}
  \tau_{\text{dec}} = t_{\text{req}} \cdot \tau_{\text{sc}}\,,
  \label{eq:dec_del}
\end{equation}
where $t_{\text{req}}$ is the required number of trials for a given codeword and $1\leq t_{\text{req}} \leq T_{\text{max}}$, i.e., the required number of decoding trials either corresponds to the number of trials until the \gls{crc} matches or to the maximum number of allowed trials.

%% file: SystemModel.tex
\section{System Model}
\label{sec:syst_mod}
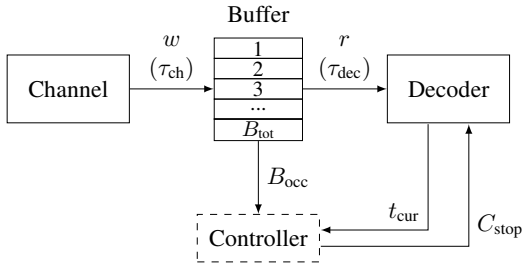
\begin{figure}[ht]
\centering%
\resizebox{0.8\columnwidth}{!}{\input{diagrams/Com_chain}}
\caption{System model containing simplified blocks of channel, buffer, controller and \gls{scf}-based decoder. Arrows indicate the data flow between the blocks.}
\label{fig:buf_mod}
\end{figure}

In this work, we use a system model where the communication chain is simplified such that parts of the transmitter, the channel and the detector, are lumped into one block denoted as the channel. \autoref{fig:buf_mod} illustrates this simplified model, where the channel acts as a data generator to the remainder of the model that is the central part of this work, i.e., the buffer, the controller, and the decoder.

In the remainder of this section, we describe the general functionality of each block of the system model, with the exception of the controller that is described at greater length in its dedicated \autoref{sec:contr_mech}. 

\subsection{Channel}
The channel block in our model acts as the generator that delivers incoming data blocks (words) to the decoder. The words are generated at a fixed time interval $\tau_{\text{ch}}$\, and stored in the buffer. The direction of the data write operation is illustrated by the arrow that is denoted by $w$ in \autoref{fig:buf_mod}. We define the channel-production interval $\tau_{\text{ch}}$ as follows:
\begin{equation}
\label{eq:chan_int}
\tau_{\text{ch}}=\upsilon_{\text{pr}}\cdot \tau_{\text{sc}}\,,
\end{equation}
where $\upsilon_{\text{pr}}\in \mathbb{R}^+$ is an additional coefficient that we call the production coefficient and $\tau_{\text{sc}}$ corresponds to the latency of one \gls{sc} decoding trial. 
The channel-production interval $\tau_{\text{ch}}$ cannot be lower than the latency of a single trial $\tau_{\text{sc}}$\,, thus $\upsilon_{\text{pr}}\geq 1$.
An increase of the production coefficient corresponds to an increase the data-production interval by the channel.

For convenience, throughout the paper, we often use the term channel-production rate, which corresponds to the inverse of the channel-production interval $\tau_{\text{ch}}$\,. 

\subsection{Buffer}
The buffer is used as memory to store words coming from the channel. The buffer is divided in slots, where each slot can accommodate one word. In this work, we consider a circular buffer. 
We denote the size of the buffer by the total number of slots $B_{\text{tot}}$, and the number of occupied slots is denoted by $B_{\text{occ}}$. One received word takes one slot in the buffer. The number of occupied slots $B_{\text{occ}}$ is provided to the controller block. 

\subsection{Decoder}
The decoder block reads the received words from the buffer. The reading event is illustrated by the arrow denoted by $r$ in \autoref{fig:buf_mod}. The decoder implements the \gls{scf} decoding algorithm, where, without loss of generality, the bit-flipping candidates are defined according to \eqref{eq:metr_dscf}. The decoder operates with a maximum number of trials $T_{\text{max}}$. However, the behavior of the block can change if the controller asserts its $C_{\text{stop}}$ signal. When $C_{\text{stop}}$ is $True$, the decoder immediately ceases the current decoding attempt, declares decoding failure, and starts processing the next word. Reading it from the buffer releases a memory slot, moving away from overflow. While $C_{\text{stop}}$ is $False$, the decoder maintains its usual behavior, i.e., attempts up to $T_{\text{max}}$ trials. The decoder provides the current number of fully applied trials $t_{\text{cur}}$ to the controller.

%% file: diagrams/Com_chain.tex
\begin{tikzpicture}[]
\node [draw, rectangle, align=center, inner sep=1pt, minimum height=1.0cm, minimum width=1.8 cm] (chan) at (0.0,0) {Channel}; 
\node[] (buf_lab) at ($(chan)+(2.8, 1.1)$){Buffer};
\node [draw, rectangle, align=center, inner sep=1pt, minimum height=1.0cm, minimum width=1.8 cm] (dec) at ($(chan)+(5.6, 0)$) {Decoder}; 

\node [draw, rectangle, align=center, inner sep=1pt, minimum height=0.3cm, minimum width=1.3 cm] (bufp1) at ($(buf_lab)+(0, -0.5)$) {\small 1}; 
\node [draw, rectangle, align=center, inner sep=1pt, minimum height=0.3cm, minimum width=1.3 cm] (bufp2) at ($(bufp1)+(0, -0.3)$) {\small 2}; 
\node [draw, rectangle, align=center, inner sep=1pt, minimum height=0.3cm, minimum width=1.3 cm] (bufp3) at ($(bufp2)+(0, -0.3)$) {\small 3}; 
\node [draw, rectangle, align=center, inner sep=1pt, minimum height=0.3cm, minimum width=1.3 cm] (bufp4) at ($(bufp3)+(0, -0.3)$) {\small ...}; 
\node [draw, rectangle, align=center, inner sep=1pt, minimum height=0.3cm, minimum width=1.3 cm] (bufp5) at ($(bufp4)+(0, -0.3)$) {\small $B_{\text{tot}}$}; 

\node [draw, rectangle, dashed, inner sep=1pt, minimum height=0.7cm, minimum width=1.8 cm] (contr) at ($(bufp5)+(0, -1.6)$){Controller};

\draw [-latex] (chan) -- node[align=center,anchor=south]{ $w$ \\ $\left(\tau_{\text{ch}}\right)$}(bufp3);
\draw [-latex] (bufp3) -- node[align=center, anchor=south]{$r$ \\ $\left(\tau_{\text{dec}}\right)$}(dec);

\draw [-latex] (bufp5) -- node[anchor=west, align=center]{$B_{\text{occ}}$}(contr);

\draw [-latex] ($(contr)+(0.92, -0.12)$) -| node[anchor=south west, align=center]{$C_{\text{stop}}$} ($(dec)+(0.3, -0.52)$);

\draw [-latex] ($(dec)+(-0.3, -0.52)$) |- node[anchor=south east, align=center]{$t_{\text{cur}}$} ($(contr)+(0.92, 0.12)$);

\end{tikzpicture}

%% file: ControlMechan.tex
\section{Control Mechanisms}
\label{sec:contr_mech}

As illustrated in \autoref{fig:buf_mod}, the controller is a key ingredient to our model. It regulates the decoder based on the number of available memory slots in the buffer. It aims to avoid buffer overflow while maximizing the error-correction performance.

During processing, the buffer has two critical states: buffer underflow and buffer overflow. Buffer underflow can easily be avoided, e.g., by suspending the decoder until the buffer is further filled with data. Furthermore, buffer underflow does not affect the error-correction performance. Buffer overflow is more challenging to deal with as it essentially requires to control the worst-case execution time of the decoder thus affecting the error-correction performance. Therefore, our work focuses on control mechanisms that cope with buffer overflow. 

In our model, the controller regulates the operation of the decoder by way of thresholds: as the number of occupied slots in the buffer gets closer to overflow, pre-defined thresholds are violated and the decoding delay is gradually restricted by lowering the maximum number of trials of the \gls{scf} decoder.

In this work, the controller can implement two different mechanisms: codeword dropping or multi-threshold. \autoref{alg:contr} illustrates the \textsc{Gen\_Ctrl\_Sigs} algorithm that generates the control signals. This algorithm covers both mechanisms that are considered. The inputs of the \textsc{Gen\_Ctrl\_Sigs} algorithm are the sets of buffer-size and trial-decoding thresholds, denoted by $\mathcal{B}$ and $\mathcal{T}$, respectively. The sets consist of multiple thresholds, where $\mathcal{B} = \{B_1,B_2, \ldots, B_P\}$ and $\mathcal{T} =  \{T_1,T_2,\ldots, T_P\}$ with $P$ being the number of thresholds in each set. The set of thresholds $\mathcal{B}$ is sorted in descending order while the set $\mathcal{T}$ is sorted in ascending order. Once sorted, each threshold from $\mathcal{B}$ corresponds to the threshold of $\mathcal{T}$ located at the same position, i.e., they form a threshold pair according to their index.  

As illustrated by \autoref{alg:contr}, the states of the buffer and of the decoder are obtained through the number of occupied buffer slots $B_{\text{occ}}$ and the current number of decoding trials $t_{\text{cur}}$. The buffer state $B_{\text{occ}}$ is compared to the elements $B_{i} \in \mathcal{B}$. When the first violation is detected, the decoder state $t_{\text{cur}}$ is compared to the threshold $T_{i} \in \mathcal{T}$ of the corresponding index $i$. If a violation is detected, the controller stops the decoder.

\begin{algorithm}
\footnotesize
\caption{Generating the controller signals based off the states of the buffer and decoder.}
\label{alg:contr}
\begin{algorithmic}[1]
\Procedure{Gen\_ Ctrl\_Sigs}{$\mathcal{B},\mathcal{T}$}
\State $B_{\text{occ}} \gets buf.getOccSlots()$, $t_{\text{cur}} \gets decoder.getCurTrials()$
\State $C_{\text{stop}} \gets False$
\For{$i$ in $1 \ldots P$}
\If{$B_{\text{occ}} > B_i$ \textbf{and} $t_{\text{cur}} \geq T_i$}
\State $C_{\text{stop}} \gets True$, \textbf{break}
\EndIf
\EndFor
\State \textbf{return} {$C_{\text{stop}}$}
\EndProcedure
\end{algorithmic}
\end{algorithm}

\subsection{Codeword-Dropping Mechanism}
\label{sec:cw_drop}

The codeword-dropping mechanism follows \autoref{alg:contr} with the single threshold pair $\{B_1,T_1\}$. Note that the threshold-violation check loop is executed only once as $P=1$. 

As will be described in \autoref{sec:multithr_thr_sel}, the codeword-dropping mechanism only comes into play when the buffer is very close to overflow, i.e., $B_1$ is almost equal to $B_{\text{tot}}$. The trial-decoding threshold is set to $T_1=0$. This way, when $B_{\text{occ}}>B_1$, the decoder is immediately stopped regardless of how many trials have been attempted, i.e., the current codeword is dropped.

\subsection{Multi-Threshold Mechanism}
\label{sec:mult_thr}

The multi-threshold mechanism follows \autoref{alg:contr} with sets of multiple thresholds. For simplicity, in this work, we propose to use sets composed of $P=3$ thresholds. The buffer-size thresholds satisfy $B_3 < B_2 < B_1 < B_{\text{tot}}$ while the trial-decoding thresholds are $T_1 < T_2 < T_3 \leq T_{\text{max}}$. The threshold pair $\{B_1, T_1\}$ is the same as codeword dropping.

To obtain the best performance and tradeoff, the number of buffer-size thresholds and their values are expected to vary depending on code length and rate, channel condition, and $T_{\text{max}}$. The general goal remains the same: evenly set the buffer-size thresholds throughout the buffer to achieve gradual control. We propose to define the trial-decoding thresholds following the methodology provided in \autoref{sec:multithr_thr_sel}.

%% file: ThrSelect.tex
\section{Threshold-Selection Methodology}
\label{sec:multithr_thr_sel}

As mentioned in the previous section, the threshold $T_1=0$, and the buffer-size thresholds $B_1,B_2,\ldots,B_P$ are evenly distributed across the buffer.
Setting $P$ to 3, only the thresholds $T_2$ and $T_3$ need to be derived.
The proposed threshold-selection methodology requires obtaining the balanced number of trials of \gls{scf} decoding from offline simulations at the channel \gls{snr} of interest and selecting the targeted production coefficient $\upsilon_{\text{pr}}$.

The key metric for determining the balanced number of trials $T_{\text{bal}}$ is the average number of decoding trials $T_{\text{av}}$ derived from offline simulations.
Experiments have shown that our system model can operate with a fixed channel-production rate without buffer overflow if the average number of trials $T_{\text{av}}$ of the decoder, restricted by $T_{\text{max}}$ alone, does not exceed the production coefficient $\upsilon_{\text{pr}}$. To establish a good tradeoff between error-correction performance and buffer-overflow prevention, we start by defining the balanced number of trials as $T_{\text{bal}}=\max(T_{\text{max}})|T_{\text{av}}<\upsilon_{\text{pr}}$. 

\begin{figure}[t]
\centering
\input{figures/Aver_Exec_Time/T_av_DSCF_2_25}
\caption{Average execution time of a \gls{scf} decoder with various maximum number of trials $T_{\text{max}}$. Two examples of production coefficients $\upsilon_{\text{pr}}$ are shown as horizontal lines.}
\label{fig:av_exec_time_dscf_2_25}
\end{figure}
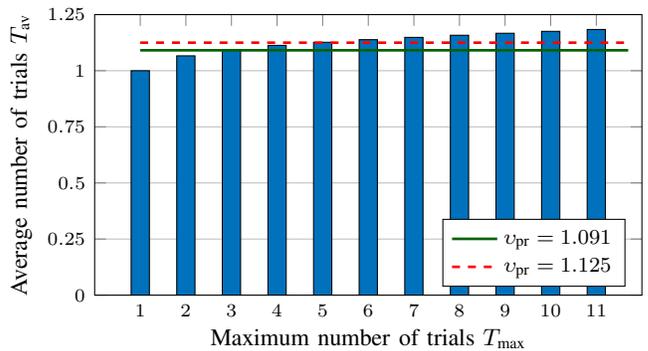

Simulations of the \gls{scf} decoder based on the setup described in \autoref{sec:results} are performed for the ideal case, i.e., $T_{\text{max}}$ is the only decoding latency restriction.
\autoref{fig:av_exec_time_dscf_2_25} shows examples of the average number of trials $T_{\text{av}}$ for various $T_{\text{max}}$ values. These results were obtained by running $10^6$ random words for each $T_{\text{max}}$ value considered and for a channel \gls{snr} of $2.25$\,dB.

To illustrate, consider the two production coefficients $\upsilon_{\text{pr}} = 1.091$ and $\upsilon_{\text{pr}} = 1.125$ represented by the horizontal lines in \autoref{fig:av_exec_time_dscf_2_25}, highlighted in solid green and dashed red, respectively.
In this example, the balanced number of trials is $T_{\text{bal}}=2$ for $\upsilon_{\text{pr}} = 1.091$ whereas it is of $4$ for $\upsilon_{\text{pr}}=1.125$.

For our proposed multi-threshold mechanism, we suggest to set thresholds $T_2$ and $T_3$ as $T_{\text{bal}}$ and $T_{\text{bal}}+1$, respectively. 
As mentioned in \autoref{sec:mult_thr}, the thresholds $B_2$ and $B_3$ are set to the middle and the head slots of the buffer. This way, the multi-threshold mechanism cuts off the high decoding trials exceeding $T_3$ once the buffer is filled up to $B_3$, and further restricts decoding to $T_2$ trials when the buffer is half full. As a further protection against buffer overflow, codeword dropping is activated when the buffer is full. The proposed methodology is applicable to other configurations, i.e., different $N$ and $k$ of the polar code, channel \gls{snr}, and $T_{\text{max}}$. 
 
We highlight that applying data rates that are too high, i.e., too low $\upsilon_{\text{pr}}$, will put too much pressure on the multi-threshold mechanism resulting in the equivalent of the codeword-dropping mechanism. Therefore, when possible, we recommend to select a data rate that results in a $T_{\text{bal}}\geq 2$.
On the other hand, if too low data rates are applied to the extent that $T_{\text{bal}}=T_{\text{max}}$, the multi-threshold mechanism is not necessary to avoid buffer overflow; $T_2=T_3=\ldots=T_P=T_{\text{max}}$. 

%% file: figures/Aver_Exec_Time/T_av_DSCF_2_25.tex
\definecolor{MyDarkGreen}{HTML}{006100}
\definecolor{MyBlue}{rgb}{0.00000,0.44700,0.74100}%
\begin{tikzpicture}
  \pgfplotsset{
    label style = {font=\fontsize{9pt}{7.2}\selectfont},
    tick label style = {font=\fontsize{7pt}{7.2}\selectfont}
  }

   \begin{axis}[%
    width=\columnwidth,
    height=\plotfigureheight\columnwidth,
    xmin=0, xmax=12,
    xtick={1,2,...,11},
    xlabel={Maximum number of trials $T_{\text{max}}$},
    xlabel style={yshift=0.4em},
    ymin=0, ymax=1.25,
    ytick={0,0.25,...,1.5},
    ylabel style={yshift=-0.1em},
    ylabel={Average number of trials $T_{\text{av}}$},
    xlabel style={yshift=-0.2em},
    ymajorgrids, 
    legend style={legend columns=1},
    legend style={legend cell align=left, align=left,font=\footnotesize},
    legend pos=south east,
    mark size=1.0pt, mark options=solid,
    ]    
    
    \addplot[ybar, bar width=0.4, fill=MyBlue, draw=black, forget plot] table[row sep=crcr]{%
      1 1.0\\
      2	1.066264\\
      3	1.094729\\
      4	1.112918\\
      5	1.126692\\
      6	1.138254\\
      7	1.148492\\
      8	1.157956\\
      9 1.166872\\
      10 1.175429\\
      11 1.183709\\
      };
    
    \addplot[thick, samples=11,line width=1.0pt,  smooth,domain=1:12,MyDarkGreen] coordinates {(1,1.091)(11.7,1.091)};

    \addplot[thick, dashed, samples=11,line width=1.0pt,  smooth,domain=1:12,red] coordinates {(1,1.125)(11.7,1.125)};    
    \legend{$\upsilon_{\text{pr}}=1.091$,$\upsilon_{\text{pr}}=1.125$} 
  
    \end{axis}

    \end{tikzpicture}%

%% file: Simul_Results.tex
\section{Simulation Results}
\label{sec:results}

We start this section with a description of our simulation methodology and continue by detailing the simulation algorithm.
The simulation results are then presented and discussed.

\subsection{Methodology}
\label{sec:sim_meth}

The simulation of the system model consists in a series of iterations with each iteration being a single unit of time. In order to represent the channel data-production interval with the production coefficient $\upsilon_{\text{pr}}$, channel and decoder blocks need to perform their operations at particular loop iterations. 
For simplicity, we normalize the time by a latency equivalent to a single \gls{sc} pass. For example, with a production coefficient $\upsilon_{\text{pr}}=1.125$ and a decoding latency $\tau_{\text{sc}}$ of $8$ units, the channel generates data every $\tau_{\text{ch}}=\upsilon_{\text{pr}}\cdot \tau_{\text{sc}}=9$ time units \eqref{eq:chan_int}. 

Before simulating our system model, we run simulations of the \gls{scf} decoder within the ideal system, i.e., with the initial maximum number of trials as the only decoding latency restriction.
To illustrate the functionality of our proposed algorithm, the random blocks of data were encoded with a $\mathcal{P}\left(1024,512\right)$ polar code and a \gls{crc} of $r=16$ bits with polynomial $z^{16}+z^{15}+z^2+1$ was used. The polar encoding algorithm is constructed for an approximate design \gls{snr} of $2.365$\,dB. Binary phase-shift keying modulation is used over an \gls{awgn} channel. Simulations were ran for $S=10^6$ random codewords at channel \glspl{snr} ranging from  $1.75$ to $2.5$\,dB. The \gls{scf} decoding algorithm with a maximum number of trials $T_{\text{max}}=11$ was used, where the bit-flipping candidates are defined according to the metric of \eqref{eq:metr_dscf}. In \cite[Eq.\,(23)]{dyn_scf}, the authors suggest adapting the constant $c$ of the metric at each \gls{snr}. Regardless, we use $c=0.3$ across all \gls{snr} values to simplify analysis. 
For each decoding word, the required number of trials is stored in the list $\psi_{\text{req}}$. 
The frame-error flag, indicating whether the word was successfully decoded or not, is stored in the list of frame-error flags $\bm{E}$. At the end of simulations, the lists $\psi_{\text{req}}$ and $\bm{E}$ are saved and used for further analysis of the system model.

Then simulations are performed for the system model of \autoref{fig:buf_mod}, using the results obtained from the simulation of the ideal system. To illustrate our algorithm, the total size of the buffer is fixed to $B_{\text{tot}}=100$ memory slots. Both codeword-dropping and multi-threshold mechanisms are simulated. For the codeword-dropping mechanism, the thresholds $B_1=99$ and $T_1=0$ are set. For the multi-threshold mechanism, the set of the buffer-size thresholds is $\mathcal{B}=\{99, 50, 10\}$. The set of corresponding trial-decoding thresholds is $\mathcal{T}=\{0, T_{\text{bal}}, T_{\text{bal}}+1\}$ and varies depending on the specific channel \gls{snr} and $\upsilon_{\text{pr}}$. For the applied configurations, the sets of three thresholds result in the optimal tradeoff between complexity and error-correction performance.

In this work, we illustrate with production coefficients that are close to the bound of 1, i.e., $\upsilon_{\text{pr}} \in \{1.091, 1.11, 1.125, 1.15, 1.2\}$ are considered. We focus on $\upsilon_{\text{pr}}$ that are close to the bound to show that the mechanism maintains a \gls{fer} near $10^{-2}$ without running into a buffer overflow even with very aggressive channel-production rates. 

For all simulations, the resulting metrics are analyzed when the buffer is filled with substantial amount of words, i.e., when the system is at the steady-state, such that the comparison is fair for different values of $\upsilon_{\text{pr}}$ and \gls{snr}.

\subsection{Simulation Algorithm}

The simulation algorithm of our system model is summarized in \autoref{alg:buf_sim}. The algorithm contains a loop, where functions corresponding to each block of the system model are called at each iteration. Each iteration of the loop corresponds to one time unit, that is used as reference to all processes in the system. The function generating the channel data is denoted by \textsc{Gen\_Data}, the function generating the controller signals is \textsc{Gen\_Ctrl\_Sigs}, and the decoder function is \textsc{Decode}. 

The functions of channel and decoder are passthrough functions with a behavior that depends on the state of their internal counters. \textsc{Gen\_Data} will add a word to the buffer after every $\tau_{\text{ch}}$ iteration loops. \textsc{Decode} will read the word from the buffer at every $\tau_{\text{dec}}=t_{\text{req}}\cdot \tau_{\text{sc}}$ iteration loops \eqref{eq:dec_del}, where the required number of trials $t_{\text{req}}$ for each decoding word $s$ is read from the list $\bm{\psi}_{\text{req}}$.

The word counter $s$ is incremented when $C_{\text{stop}}$ is raised, i.e., when either one of the thresholds is violated or when the decoder completed decoding according to $t_{\text{req}}$. At the same condition, the final current number of trials is saved to the list of resulting number of trials $\bm{\psi}_{\text{res}}$. Simulation ends when all $S$ decoding words are processed. The number of occupied buffer slots is stored in the list $\bm{\chi}_{\text{occ}}$ at every loop iteration. 

At the end of simulation, \textsc{Calc\_fer\_Impact} calculates the binary list of resulting frame-error flags $\bm{E}^{\prime}$ indicating which words were successfully decoded and which were not. This list differs from the list of original frame-error flags $\bm{E}$ obtained from the simulation of the ideal system.
A decoding error is declared when the ideal system failed to decode the word or when there is an early decoder stoppage ($\psi_{\text{res}}(s)<\psi_{\text{req}}(s)$).

\begin{algorithm}[t]
\caption{Simulation algorithm of the system model with the fixed channel-generated data rate.}
\label{alg:buf_sim}
\begin{algorithmic}[1]\footnotesize
\Inputs{$S$, $B_{\text{tot}}$, $\tau_{\text{ch}}$, $\tau_{\text{sc}}$, $\psi_{\text{req}}$, $\bm{E}$, $\mathcal{B}$,  $\mathcal{T}$}
\Procedure{Sim\_Syst\_Model}{}
\State $\bm{\psi}_{\text{res}} \gets \{0,0,\ldots 0\}$, $\bm{\chi}_{\text{occ}}\gets \{0,0,\ldots 0\}$
\State $buf \gets \textsc{Create\_Buf}\left( B_{\text{tot}}\right)$
\State $t_{\text{req}}\gets \bm{\psi}_{\text{req}}(1)$, $s \gets 1$, $i\gets 1$
\While{$s \neq S$}
\State $\textsc{Gen\_Data}\left(buf, \tau_{\text{ch}} \right)$
\State $C_{\text{stop}}\gets   \textsc{Gen\_Ctrl\_Sigs}\left(\mathcal{B}, \mathcal{T}, t_{\text{req}} \right)$
\State $t_{\text{cur}} \gets \textsc{Decode}\left(buf,C_{\text{stop}}, t_{\text{req}}, \tau_{\text{sc}}\right)$
\If{$(C_{\text{stop}}==True)$}
\State $\bm{\psi_{\text{res}}}(s) \gets t_{\text{cur}}$, $s\gets s+1$, $t_{\text{req}}\gets \bm{\psi}_{\text{req}}(s)$
\EndIf
\State $\bm{\chi}_{\text{occ}}\left( i\right)\gets buf.B_{\text{occ}}$
\State $i++$
\EndWhile
\State $\bm{E}^{\prime} \gets  \textsc{Calc\_fer\_impact}\left(\bm{\psi}_{\text{res}},\bm{\psi}_{\text{req}},\bm{E} \right)$
\State \textbf{return} $\left(\bm{\chi}_{\text{occ}}, \bm{E}^{\prime} \right)$
\EndProcedure
\end{algorithmic}
\end{algorithm}

\subsection{State of the Buffer Over the Course of Simulation}
\label{sec:buf_slots_analys}

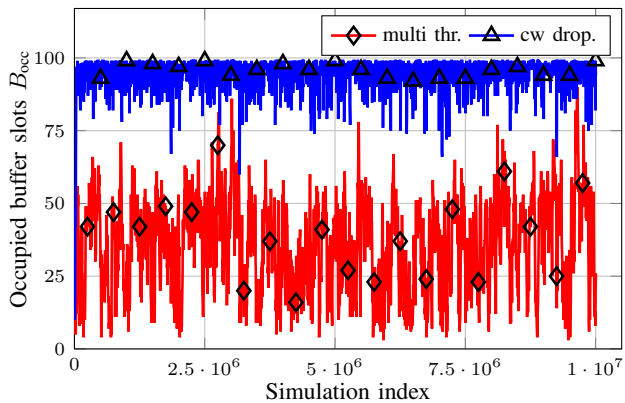
\begin{figure}[t]
\centering
\input{figures/Buf_slots_DSCF/Used_slots_200_2_25_9_8}
\caption{Number of occupied buffer slots over the course of a simulation of the codeword-dropping and the multi-threshold mechanisms for \gls{snr} of $2.25$\,dB and $\upsilon_{\text{pr}}=1.125$.
}
\label{fig:used_slots_dscf_1_125}
\end{figure}

\autoref{fig:used_slots_dscf_1_125} shows the number of used buffer slots over the course of simulation, where the words come from the channel at a fixed rate that corresponds to a production coefficient $\upsilon_{\text{pr}}=1.125$ and the channel \gls{snr} is of $2.25$\,dB. The codeword-dropping mechanism is depicted in blue while the multi-threshold is in red. From the figure, we can see that both mechanisms effectively prevent buffer overflow, i.e., buffer occupied slots never reach $B_{\text{tot}}=100$ slots.

\subsection{Error-correction performance}

\autoref{fig:fer_dscf_var_snr_9_8} shows the \gls{fer} of the model, where the controller implements the codeword-dropping (blue) and multi-threshold mechanisms (red). 
Simulations are for various \glspl{snr}, but for a fixed channel-production rate corresponding to $\upsilon_{\text{pr}}=1.125$. As such, the channel-production interval is close to the delay of a single \gls{scf} trial.
The black curve is the ideal performance provided for reference.
The figure shows that, at low channel \gls{snr}, both considered control mechanisms experience a degradation of the error-correction performance compared to the ideal case. This gap is reduced as the channel improves; the loss is virtually nonexistent at a \gls{snr} of $2.375$\,dB. Across the range, we see that the multi-threshold mechanism either matches or outperforms the codeword-dropping mechanism. At the point of interest for wireless communication, a \gls{fer} of $10^{-2}$ is achieved by the \gls{scf} decoder within the ideal system at approximately $2.25$\,dB. The codeword-dropping and the multi-threshold mechanisms show performance losses of approximately $0.1$\,dB and $0.0625$\,dB respectively.

\begin{figure}[t]
\input{figures/FER_DSCF/FER_DSCF_2_25_var_rate}\vspace{-5pt}
\caption{\gls{fer} of the codeword-dropping and the multi-threshold mechanisms for the range of the $\upsilon_{\text{pr}}$ and \gls{snr} of $2.25$\,dB.
}
\label{fig:fer_dscf_2_25_var_rate}
\end{figure}
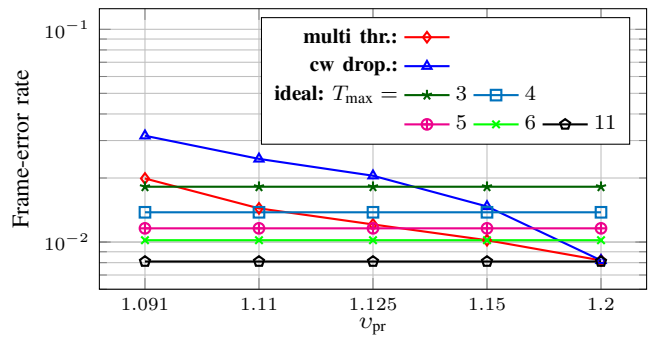

\autoref{fig:fer_dscf_2_25_var_rate} also shows the \gls{fer} of the model for both mechanisms, but for a fixed \gls{snr} of $2.25$\,dB and various $\upsilon_{\text{pr}}$\,.
\gls{scf} with $T_{\text{max}}=11$ is applied for both mechanisms.
Although it cannot be sustained, the ideal performance for various $T_{\text{max}}$ values are shown as horizontal lines for reference. From the figure, it can be seen that at lower production coefficients both codeword-dropping and multi-threshold mechanisms have a loss in error-correction performance compared to the ideal case with $T_{\text{max}}=11$. At $\upsilon_{\text{pr}}=1.091$, the \gls{fer} is even worse than the ideal case with $T_{\text{max}}=3$. The gap reduces as the production coefficient increases. The multi-threshold mechanisms fares better than codeword dropping across the whole range. At $\upsilon_{\text{pr}}=1.125$, the \gls{fer} of the multi-threshold mechanism reaches the ideal case for $T_{\text{max}}=5$. Both mechanisms match the ideal \gls{fer} for $T_{\text{max}}=11$ at $\upsilon_{\text{pr}}=1.2$. 

\begin{figure}[t]
\input{figures/FER_DSCF/FER_DSCF_2_25_var_snr}\vspace{-5pt}
\caption{\gls{fer} of the codeword-dropping and the multi-threshold mechanisms for the range of \gls{snr} and $\upsilon_{\text{pr}}=1.125$. 
}
\label{fig:fer_dscf_var_snr_9_8}
\end{figure}
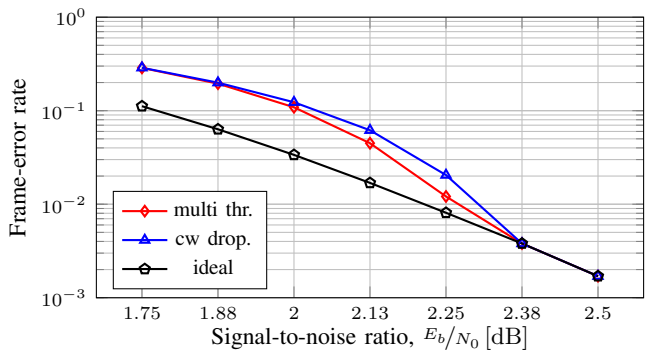

%% file: figures/Buf_slots_DSCF/Used_slots_200_2_25_9_8.tex
\definecolor{MyDarkGreen}{HTML}{006100}

  \pgfplotsset{
 label style = {font=\fontsize{9pt}{7.2}\selectfont},
 tick label style = {font=\fontsize{7pt}{7.2}\selectfont}
  }

\begin{tikzpicture}
  \begin{axis}[%
 width=\columnwidth,
 height=1.15*\plotfigureheight\columnwidth,
 xmin=0, xmax=2100,
 xlabel={Simulation index},
 xlabel style={yshift=0.4em},
 xticklabel={\pgfkeys{/pgf/fpu}\pgfmathparse{5000*\tick}$\pgfmathprintnumber{\pgfmathresult}$},
 ymin=0, ymax=117,
 ytick={0,25,...,125},
 ylabel style={yshift=-0.6em},
 ylabel={Occupied buffer slots $B_{\text{occ}}$},
 xmajorgrids, ymajorgrids,
 yminorticks, yminorgrids,
 legend style={legend cell align=left, align=left,font=\footnotesize},
 legend columns=2,
 mark size=3.0pt, mark options=solid,
 ]   
  \addplot [color=red, line width=1.0pt, mark=diamond, mark options={color=black,scale=1.1}, mark repeat=100, mark phase=50]
 table[x=idx,y=slots]{data/used_slots_dscf_100_multithr_2_25_9_8.tex};
  \addlegendentry{multi thr.}
   
  \addplot [color=blue, line width=1.0pt, mark=triangle, mark options={color=black,scale=1.1}, mark repeat=100, mark phase=100]
 table[x=idx,y=slots]{data/used_slots_dscf_100_cw_drop_2_25_9_8.tex};
  \addlegendentry{cw drop.}

   \end{axis}
\end{tikzpicture}%

%% file: figures/FER_DSCF/FER_DSCF_2_25_var_snr.tex
\begin{tikzpicture}
  \pgfplotsset{
    label style = {font=\fontsize{9pt}{7.2}\selectfont},
    tick label style = {font=\fontsize{7pt}{7.2}\selectfont}
  }

  \begin{semilogyaxis}[%
    width=\columnwidth,
    height=\plotfigureheight\columnwidth,
    xtick={1.75,1.875,...,2.5},
    xlabel={Signal-to-noise ratio, $\nicefrac{E_b}{N_0} \left[ \mathrm{dB} \right]$},
    xlabel style={yshift=0.4em},
    ymin=1e-3, ymax=1,
    ylabel style={yshift=-0.1em},
    ylabel={Frame-error rate},
    yminorticks, xmajorgrids,
    ymajorgrids, yminorgrids,
    legend pos=south west,
    legend style={font=\footnotesize},
    mark size=1.0pt, mark options=solid,
    ]     
    
    \addplot[color=red, mark size=2pt, mark=diamond, line width=0.8pt]
    table[row sep=crcr]{%
      1.75	0.2867\\
      1.875	0.1946\\
      2.0	0.1088\\
      2.125	0.0449\\
      2.25	0.0121\\
      2.375	0.0038\\
      2.5	0.0017\\
      };
    \addlegendentry{multi thr.} 
    
    \addplot[color=blue, mark size=2pt, mark=triangle, line width=0.8pt]
    table[row sep=crcr]{%
      1.75	0.2879\\
      1.875	0.1994\\
      2.0	0.1231\\
      2.125	0.0618\\
      2.25	0.0205\\
      2.375	0.0038\\
      2.5	0.0017\\
      };
    \addlegendentry{cw drop.}  
       
    \addplot[color=black, mark size=2pt, mark=pentagon, line width=0.8pt]
    table[row sep=crcr]{%
      1.75	0.1117\\
      1.875	0.0633\\      
      2.0	0.0337\\      
      2.125	0.0169\\        
      2.25	0.0081\\      
      2.375	0.0038\\
      2.5	0.0017\\
      };
    \addlegendentry{ideal}    
     
  \end{semilogyaxis}

\end{tikzpicture}%

%% file: figures/FER_DSCF/FER_DSCF_2_25_var_rate.tex
\definecolor{MyDarkGreen}{HTML}{006100}

\begin{tikzpicture}
  \pgfplotsset{
    label style = {font=\fontsize{9pt}{7.2}\selectfont},
    tick label style = {font=\fontsize{7pt}{7.2}\selectfont}
  }
   \begin{semilogyaxis}[%
    width=\columnwidth,
    height=\plotfigureheight\columnwidth,
    xtick=data,
    major x tick style = transparent,
    symbolic x coords={{$1.091$}, {$1.11$},{$1.125$},{$1.15$},{$1.2$}},
    xlabel={$\upsilon_{\text{pr}}$},
    xlabel style={yshift=0.4em},
    ymin=6e-3, ymax=1.25e-1,
    ylabel style={yshift=-0.1em},
    ylabel={Frame-error rate},
    yminorticks, xmajorgrids,
    ymajorgrids, yminorgrids,
    legend style={legend columns=4, font=\footnotesize, column sep=0mm},
    legend pos=north east,
    mark size=1.0pt, mark options=solid,
    ]

    \addlegendimage{empty legend}
    \addlegendentry[anchor=east]{\textbf{multi thr.:}}

    \addplot[color=red, mark size=2pt, mark=diamond, line width=0.8pt] coordinates{({$1.091$}, 0.0199) ({$1.11$}, 0.0144) ({$1.125$}, 0.0121) ({$1.15$}, 0.0102) ({$1.2$}, 0.0082)};
    \addlegendentry{}
    \addlegendimage{empty legend}\addlegendentry{}
    \addlegendimage{empty legend}\addlegendentry{}

    \addlegendimage{empty legend}
    \addlegendentry[anchor=east]{\textbf{cw drop.:}}
    \addplot[color=blue, mark size=2pt, mark=triangle, line width=0.8pt] coordinates{({$1.091$}, 0.0316)  ({$1.11$},0.0246) ({$1.125$}, 0.0205) ({$1.15$}, 0.0147) ({$1.2$}, 0.0082)};
    \addlegendentry{}
    \addlegendimage{empty legend}\addlegendentry{}
    \addlegendimage{empty legend}\addlegendentry{}

    \addlegendimage{empty legend}
    \addlegendentry[anchor=east]{\textbf{ideal:} $T_{\text{max}}=$}
    \addplot[color=MyDarkGreen, mark size=2pt, mark=star, line width=0.8pt] coordinates{({$1.091$}, 0.0182) ({$1.11$},0.0182) ({$1.125$}, 0.0182) ({$1.15$}, 0.0182) ({$1.2$}, 0.0182)};
    \addlegendentry{$3$}
    
    \addplot[color=MyBlue, mark size=2pt, mark=square, line width=0.8pt] coordinates{({$1.091$}, 0.0138) ({$1.11$},0.0138) ({$1.125$}, 0.0138) ({$1.15$}, 0.0138) ({$1.2$}, 0.0138)};
    \addlegendentry{$4$}
    \addlegendimage{empty legend}\addlegendentry{}
    
    \addlegendimage{empty legend}
    \addlegendentry[anchor=east]{\phantom{\textbf{ideal:}}}
    \addplot[color=magenta, mark size=2pt, mark=oplus, line width=0.8pt] coordinates{({$1.091$}, 0.0116) ({$1.11$},0.0116) ({$1.125$}, 0.0116) ({$1.15$}, 0.0116) ({$1.2$}, 0.0116)};
    \addlegendentry{$5$}
    
    \addplot[color=green, mark size=2pt, mark=x, line width=0.8pt] coordinates{({$1.091$}, 0.0102) ({$1.11$},0.0102) ({$1.125$}, 0.0102) ({$1.15$}, 0.0102) ({$1.2$}, 0.0102)};
    \addlegendentry{$6$}

    \addplot[color=black, mark size=2pt, mark=pentagon, line width=0.8pt] coordinates{({$1.091$}, 0.0081) ({$1.11$},0.0081) ({$1.125$}, 0.0081) ({$1.15$}, 0.0081) ({$1.2$}, 0.0081)};
    \addlegendentry{$11$}

  \end{semilogyaxis}

\end{tikzpicture}%

%% file: Conclusion.tex
\section{Conclusion}
\label{sec:conclusion}
In this work, we proposed a control algorithm that adjusts the execution time of a \gls{scf}-based decoder in realtime, allowing it to sustain operation without buffer overflow with a channel that produces data with a fixed rate that approaches that of a single decoding trial. By using multiple thresholds, the proposed mechanism is shown to allow an \gls{scf}-based decoder to operate in a system with a fixed channel-production rate that is $1.125$ times lower than the rate associated to a single decoding trial while preventing buffer overflow. In the region of interest for wireless communications, this at the cost of a small error-correction performance of approximately $0.0625$\,dB in comparison to the ideal but unsustainable case.